\documentstyle[12pt,prc,aps,psfig]{revtex}
\tighten
\begin{document}
\draft

\title{More on nucleon-nucleon cross sections in symmetric and \\
asymmetric matter}
\author{F. Sammarruca and P. Krastev}
\address{Physics Department, University of Idaho, Moscow, ID 83844, U.S.A}
\date{\today}
\maketitle
\begin{abstract}
Following a recent work, we present numerical results for total 
two-nucleon effective cross sections in isospin symmetric and asymmetric matter.
The present calculations include the additional effect of Pauli
blocking of the final state. 

\end{abstract}
\narrowtext

\section{Brief description of the updated calculation} 

In a recent paper \cite{SK05}, we discussed in-medium 
nucleon-nucleon (NN) cross sections in a dense hadronic 
environment with and without isospin asymmetry.                                   
This note is a continuation of our previous work and the present purpose is to provide a large set
of numerical results for possible applications in 
transport equations.                                                              
 
The cross sections we display here differ from those of Ref.~\cite{SK05} 
due to the inclusion of Pauli blocking of the final state.
The need for such correction may be dependent on the details of 
the chosen Boltzmann-Uehling-Uhlenbeck (BUU) equation, as this effect is sometimes
incorporated at the level of the transport calculation. In such case, though, 
a Pauli blocking function is usually applied together with in-medium {\it differential} cross sections.
However, 
{\it total} effective cross sections with properly Pauli-blocked
final states should be a better indication of the degree of suppression  the 
interaction actually undergoes in the medium. Furthermore, this correction must be
included for a realistic calculation of the nucleon mean free path in nuclear 
matter. 

We calculate the total effective cross section as 
\begin{equation}
\sigma(q_0,P_{tot},\rho)= \int \frac{d\sigma}{d\Omega}                               
Q(q_0,P_{tot},\theta, \rho)d\Omega , 
\end{equation}
where
$\frac{d\sigma}{d\Omega} $                                    
is given by the usual sum of amplitudes squared and phase space factors and      
$Q$ is the Pauli operator.              
Ignoring Pauli blocking on the final momenta amounts to setting $Q$=1 in the integrand above, 
as done in Ref.~\cite{SK05}. 
(Note that the intermediate states are always Pauli-blocked in the microscopic calculation 
of the amplitudes.)                                                                
All definitions as in Ref.\cite{SK05}, that is, $q_0$ 
is the nucleon momentum
in the two-body c.m. frame                   
(or the relative momentum of the two nucleons), and $P_{tot}$ is the total momentum 
of the two-nucleon system with respect to the nuclear matter rest frame. 
In free space, the momentum $q_0$ and the incident laboratory kinetic energy
are related through the well-known
formula $T_{lab} = 2q_0^2/m$, with $m$ the free nucleon mass.       

The presence of the Pauli operator in Eq.(1) restricts the integration domain
to \cite{SM} 
\begin{equation}
\frac{k_F^2-P^2-q_0^2}{2Pq_0} \leq  \cos \theta \leq
\frac{P^2+q_0^2-k_F^2}{2Pq_0},                                 
\end{equation}
where $k_F$ is the Fermi momentum.
The integral becomes zero if the upper limit is negative, whereas 
the full angular range is used if the upper limit is greater than one.
(Following a previously used convention \cite{SM}, $P$ in Eq.(2) denotes        
one-half the total momentum.) Notice that the angle $\theta$ in Eq.(2), namely the 
angle between the directions of $\vec{q_0}$ and $\vec{P}$, is also the colatitude 
of $\vec{q_0}$ in a system where the $z$-axis is along 
$\vec{P}$ and, thus, in such system it coincides with the scattering angle to be 
integrated over in Eq.(1). 
            
Except for the inclusion of $Q$ in Eq.(1), 
the present microscopic calculation is the same as in Ref.\cite{SK05}.
We adopt the Bonn-B potential and the Dirac-Brueckner-Hartree-Fock (DBHF)
approach to nuclear matter, unless otherwise stated.
For simplicity, we use in-vacuum kinematics to define the total two-nucleon momentum
in the nuclear matter rest frame (that is, the target nucleon is, on the average,
at rest).

\section{Numerical results for $pp$ and $np$ effective cross sections in       
symmetric matter} 

We begin with discussing our predictions in symmetric matter. These are
displayed in Table I and II for $pp$ and $np$ scattering, respectively.
The given values of $q_0$ cover a range between approximately 20 MeV and 350 MeV
in terms of in-vacuum laboratory kinetic energy. 
The range of densities goes from zero to about twice saturation density.
Due to the presence of the Pauli operator in Eq.(1), it's clear that the 
cross sections will drop to zero at certain densities depending on the 
value of the momentum. \\ \\ 
\noindent
{\bf $pp$ cross sections:}\\ \\ 
At very low energy, 
we observe strong sensitivity to variations of the Fermi momentum when                
approaching the density where the cross
section becomes identically zero.                            

At very low (fixed) density, the cross section decreases with 
energy, 
a behavior similar to what happens in free space.
At higher densities, though, the cross section grows with                    
energy. This is due to the fact that the Pauli operator in Eq.(1) becomes less effective at the 
higher energies.

For fixed momentum, the cross section typically decreases with increasing
density. 
In Ref.~\cite{SK05} we discussed the tendency (of the $pp$ cross section, in particular)
to rise again at high density and for the higher momenta.                               
This typical 
``Dirac effect'', already observed in previous DBHF calculations, is of course
still present in the interaction, but is less noticeable in these cross sections
due to the additional Pauli suppression in Eq.(1). To better demonstrate its origin,
we also show results obtained with the conventional 
Brueckner-Hartree-Fock approach (BHF) (Table III and IV for $pp$ and $np$, 
respectively). 
The differences are most apparent when moving down to the higher densities 
in the last few columns of Table I and comparing with the corresponding entries
in Table III.         \\ \\     
\noindent
{\bf $np$ cross sections:}\\ \\ 
As in the $pp$ case,                                          
very strong $k_F$-sensitivity can be seen at the lowest momenta just before the 
cross section is totally suppressed. 

For fixed momentum, the cross section typically decreases with increasing
density. 
An interesting difference between $pp$ and $np$ cross sections is the fact that the latter
rises with density at very low density and for the lowest momenta.                          
We traced this effect to an enhancement of the $^{3}S_{1}$ partial wave and 
did not observe a similar phenomenon in isospin-1 partial waves. 
The presence of low-density ``peaks'' (in the $np$ channel in particular), was 
discussed before as a possible bound-state signature \cite{hh1,hh2}. 

Finally,  
the differences between the values in Table II and Table IV for high momentum
and high density are much less pronounced than in the $pp$ case. 

\section{Numerical results for $pp$ and $nn$ effective cross sections in 
asymmetric matter}

Predictions for $pp$ and $nn$ total cross sections in isospin-asymmetric
matter are displayed in Table V for different degrees of asymmetry. 
These are obtained with our DBHF model. 
The asymmetry parameter, $\alpha$, is defined in the usual way in terms of 
neutron and proton densities, namely, 
$\alpha$=$\frac{\rho_n - \rho _p}{\rho _n + \rho _p}$.                               
We ignore the $\alpha$-dependence in the $np$ channel as we 
previously found it to be very weak (due to the ``competing'' roles of 
protons and neutrons). 

The $nn$ cross section is (almost always) smaller than the $pp$ cross section.
This is due to the                                                              
additional Pauli blocking included in Eq.(1), together with the fact that the 
neutron's 
Fermi momentum is larger than the proton's. In fact, as a consequence of       
Eq.(1), the $nn$ effective cross
section drops to zero more quickly (for the same average density). This is
apparent from the Table. Although at low density the $nn$ cross section tends to
start larger, Pauli blocking soon takes over. 
Particularly for large values of $\alpha$, the $pp$ cross section
``survives'' much larger densities than the $nn$. This is to be expected,       
because the proton Fermi momentum, $k_F^p$=$k_F(1-\alpha)^{1/3}$,              
tends to remain small for large $\alpha$.

All of the above
suggests the following observation: the region of the density/momentum phase space       
where $\sigma_{nn}$ is nearly or entirely suppressed whereas $\sigma_{pp}$ is 
still considerably different than zero should be a suitable ground to 
look for a clear experimental signature of their difference.

Some comments are in place concerning the relative sizes of $\sigma_{pp}$
and $\sigma_{nn}$.
Predictions based on scaling the free-space cross section through the use of  
effective masses in the phase-space factor will generally have the trend:
\begin{itemize}
\item 
 $\sigma_{nn}> \sigma_{pp}$, if $m^{*}_n>m^{*}_p $\cite{Li05};
\item
 $\sigma_{pp}> \sigma_{nn}$, if $m^{*}_p>m^{*}_n $\cite{QLi04}. 
\end{itemize}
Notice that our predictions in Ref.~\cite{SK05} are qualitatively consistent
with the first case above, due to the dominant role of the effective masses
in that calculation (we predict $m^{*}_n>m^{*}_p$ \cite{SBK}).

Obviously, empirical constraints that might shed light on these issues
can only be indirect, for instance through analyses of carefully selected heavy-ion
observables. Such efforts are presently under way \cite{MSU,Li05}. In fact, a recent study
indicates that the transverse flow of neutrons and protons may be a reliable probe
of in-medium cross section \cite{Li05}. 
As the present discussion suggests, 
when comparing theoretical predictions with experimental constraints                           
it is important to be clear about the nature 
of the extracted effective cross section, namely what medium effects are
included in the quantity that is being constrained by the analysis.
\\ \\ 
\begin{center}
{\bf ACKNOWLEDGMENTS}
\end{center}
The authors        
acknowledge
financial support from the U.S. Department of Energy under grant number DE-FG02-03ER41270.

\newpage 

\begin{table}                
\centering \caption                                                    
{$pp$ total effective cross sections in symmetric matter 
calculated with the DBHF model and according to Eq.(1).                             
Kinematics and definitions of variables are explained in the text. 
} 
\vspace{5mm}

\begin{tabular}{|c|c|c|c|c|c|c|c|}
\hline

$k_{F}(fm^{-1})$ & $\sigma _{pp}(mb)$ & $\sigma _{pp}(mb)$ &
                   $\sigma _{pp}(mb)$ & $\sigma _{pp}(mb)$ &
                   $\sigma _{pp}(mb)$ & $\sigma _{pp}(mb)$ &
                   $\sigma _{pp}(mb)$\\
                &\scriptsize{$q_0=100MeV$}&
                 \scriptsize{$q_0=150MeV$}&
                 \scriptsize{$q_0=200MeV$}&
                 \scriptsize{$q_0=250MeV$}&
                 \scriptsize{$q_0=300MeV$}&
                 \scriptsize{$q_0=350MeV$}&
                 \scriptsize{$q_0=400MeV$}\\
\hline
0.0 & 171.2 & 69.33 & 39.86 & 28.93 & 23.96 & 21.47 & 20.27\\
0.2 & 157.1  & 61.71 & 34.97 & 25.56 & 21.70 & 20.05 & 19.47\\
0.4 & 133.4  & 52.91 & 29.90 & 22.32 & 19.63 & 18.80 & 18.82\\
0.6 &  64.34 & 42.58 & 24.97 & 19.39 & 17.85 & 17.78 & 18.31\\
0.8 &   0.00 & 29.06 & 20.11 & 16.71 & 16.30 & 16.92 & 17.91\\
0.9 &   0.00 & 19.86 & 17.61 & 15.42 & 15.57 & 16.52 & 17.71\\
1.0 &   0.00 & 8.980 & 14.99 & 14.14 & 14.86 & 16.13 & 17.51\\
1.1 &   0.00 &  0.00 & 11.98 & 12.70 & 14.07 & 15.72 & 17.31\\
1.2 &   0.00 &  0.00 & 8.602 & 11.14 & 13.24 & 15.29 & 17.13\\
1.3 &   0.00 &  0.00 & 5.025 & 9.478 & 12.37 & 14.89 & 17.01\\
1.4 &   0.00 &  0.00 & 1.583 & 7.793 & 11.54 & 14.56 & 16.98\\
1.5 &   0.00 &  0.00 &  0.00 & 6.123 & 10.77 & 14.34 & 17.07\\
1.6 &   0.00 &  0.00 &  0.00 & 4.466 & 10.05 & 14.20 & 17.25\\
1.7 &   0.00 &  0.00 &  0.00 & 2.698 & 9.286 & 14.04 & 17.40\\
\hline

\end{tabular}
\end{table}

\begin{table}               
\centering \caption                                      
{$np$ total effective cross sections in symmetric matter 
calculated with the DBHF model and according to Eq.(1).                                 
Kinematics and definitions of variables as explained in the text. 
}                                               

\vspace{5mm}

\begin{tabular}{|c|c|c|c|c|c|c|c|}
\hline

$k_{F}(fm^{-1})$ & $\sigma _{np}(mb)$ & $\sigma _{np}(mb)$ &
                   $\sigma _{np}(mb)$ & $\sigma _{np}(mb)$ &
                   $\sigma _{np}(mb)$ & $\sigma _{np}(mb)$ &
                   $\sigma _{np}(mb)$\\
                &\scriptsize{$q_0=100MeV$}&
                 \scriptsize{$q_0=150MeV$}&
                 \scriptsize{$q_0=200MeV$}&
                 \scriptsize{$q_0=250MeV$}&
                 \scriptsize{$q_0=300MeV$}&
                 \scriptsize{$q_0=350MeV$}&
                 \scriptsize{$q_0=400MeV$}\\
\hline
0.0 & 453.3  & 174.1  & 86.41 & 55.02 & 41.65 & 34.60 & 30.15\\
0.2 & 459.7  & 159.6  & 75.48 & 47.35 & 36.23 & 30.78 & 27.52\\
0.4 & 508.3  & 144.7  & 63.55 & 39.26 & 30.68 & 26.85 & 24.72\\
0.6 & 421.6  & 128.8  & 52.49 & 31.98 & 25.73 & 23.40 & 22.26\\
0.8 &   0.00 & 101.3  & 42.49 & 25.71 & 21.51 & 20.45 & 20.18\\
0.9 &   0.00 &  72.45 & 37.58 & 22.91 & 19.61 & 19.14 & 19.23\\
1.0 &   0.00 &  31.84 & 32.37 & 20.28 & 17.85 & 17.89 & 18.35\\
1.1 &   0.00 &   0.00 & 25.82 & 17.45 & 15.96 & 16.56 & 17.39\\
1.2 &   0.00 &   0.00 & 17.91 & 14.50 & 14.06 & 15.19 & 16.39\\
1.3 &   0.00 &   0.00 & 9.578 & 11.44 & 12.14 & 13.81 & 15.39\\
1.4 &   0.00 &   0.00 & 2.628 & 8.512 & 10.35 & 12.52 & 14.48\\
1.5 &   0.00 &   0.00 &  0.00 & 5.899 & 8.763 & 11.41 & 13.69\\
1.6 &   0.00 &   0.00 &  0.00 & 3.761 & 7.454 & 10.52 & 13.08\\
1.7 &   0.00 &   0.00 &  0.00 & 2.018 & 6.386 & 9.810 & 12.60\\
\hline

\end{tabular}
\end{table}
\begin{table}                
\centering \caption                                                    
{As in Table I, but with the BHF model.                      
} 
\vspace{5mm}

\begin{tabular}{|c|c|c|c|c|c|c|c|}
\hline

$k_{F}(fm^{-1})$ & $\sigma _{pp}(mb)$ & $\sigma _{pp}(mb)$ &
                   $\sigma _{pp}(mb)$ & $\sigma _{pp}(mb)$ &
                   $\sigma _{pp}(mb)$ & $\sigma _{pp}(mb)$ &
                   $\sigma _{pp}(mb)$\\
                &\scriptsize{$q_0=100MeV$}&
                 \scriptsize{$q_0=150MeV$}&
                 \scriptsize{$q_0=200MeV$}&
                 \scriptsize{$q_0=250MeV$}&
                 \scriptsize{$q_0=300MeV$}&
                 \scriptsize{$q_0=350MeV$}&
                 \scriptsize{$q_0=400MeV$}\\
\hline
0.0 & 171.2 & 69.33 & 39.86 & 28.93 & 23.96 & 21.47 & 20.27\\
0.2 & 156.8  & 62.36 & 35.68 & 26.06 & 21.90 & 19.95 & 19.12\\
0.4 & 131.4  & 53.62 & 30.85 & 22.91 & 19.69 & 18.33 & 17.86\\
0.6 &  61.28 & 42.74 & 25.72 & 19.71 & 17.49 & 16.72 & 16.61\\
0.8 &   0.00 & 28.39 & 20.36 & 16.53 & 15.33 & 15.15 & 15.39\\
0.9 &   0.00 & 19.02 & 17.57 & 14.95 & 14.27 & 14.37 & 14.78\\
1.0 &   0.00 & 8.431 & 14.68 & 13.36 & 13.21 & 13.60 & 14.17\\
1.1 &   0.00 &  0.00 & 11.72 & 11.83 & 12.21 & 12.87 & 13.60\\
1.2 &   0.00 &  0.00 & 8.404 & 10.15 & 11.08 & 12.04 & 12.93\\
1.3 &   0.00 &  0.00 & 4.900 & 8.353 & 9.831 & 11.08 & 12.15\\
1.4 &   0.00 &  0.00 & 1.526 & 6.526 & 8.526 & 10.06 & 11.30\\
1.5 &   0.00 &  0.00 &  0.00 & 4.714 & 7.175 & 8.973 & 10.37\\
1.6 &   0.00 &  0.00 &  0.00 & 3.011 & 5.823 & 7.836 & 9.380\\
1.7 &   0.00 &  0.00 &  0.00 & 1.493 & 4.499 & 6.665 & 8.332\\ 
\hline

\end{tabular}
\end{table}
\begin{table}                
\centering \caption                                                    
{As in Table II, but with the BHF model.                      
} 
\vspace{5mm}

\begin{tabular}{|c|c|c|c|c|c|c|c|}
\hline

$k_{F}(fm^{-1})$ & $\sigma _{np}(mb)$ & $\sigma _{np}(mb)$ &
                   $\sigma _{np}(mb)$ & $\sigma _{np}(mb)$ &
                   $\sigma _{np}(mb)$ & $\sigma _{np}(mb)$ &
                   $\sigma _{np}(mb)$\\
                &\scriptsize{$q_0=100MeV$}&
                 \scriptsize{$q_0=150MeV$}&
                 \scriptsize{$q_0=200MeV$}&
                 \scriptsize{$q_0=250MeV$}&
                 \scriptsize{$q_0=300MeV$}&
                 \scriptsize{$q_0=350MeV$}&
                 \scriptsize{$q_0=400MeV$}\\
\hline
0.0 & 453.3  & 174.1  & 86.41 & 55.02 & 41.65 & 34.60 & 30.15\\
0.2 & 462.3  & 162.0  & 77.26 & 48.79 & 37.38 & 31.61 & 28.04\\
0.4 & 520.4  & 149.4  & 66.39 & 41.42 & 32.38 & 28.06 & 25.39\\
0.6 & 464.9  & 135.6  & 55.71 & 34.23 & 27.46 & 24.57 & 22.78\\
0.8 &   0.00 & 109.2  & 45.62 & 27.64 & 22.91 & 21.30 & 20.34\\
0.9 &   0.00 &  78.78 & 40.53 & 24.59 & 20.77 & 19.76 & 19.18\\
1.0 &   0.00 &  34.49 & 35.03 & 21.68 & 18.72 & 18.26 & 18.05\\
1.1 &   0.00 &   0.00 & 28.91 & 19.00 & 16.85 & 16.89 & 17.02\\
1.2 &   0.00 &   0.00 & 20.99 & 16.15 & 14.89 & 15.43 & 15.89\\
1.3 &   0.00 &   0.00 & 11.91 & 13.16 & 12.88 & 13.88 & 14.66\\
1.4 &   0.00 &   0.00 & 3.466 & 10.13 & 10.90 & 12.33 & 13.42\\
1.5 &   0.00 &   0.00 &  0.00 & 7.161 & 8.971 & 10.78 & 12.14\\
1.6 &   0.00 &   0.00 &  0.00 & 4.442 & 7.137 & 9.263 & 10.85\\
1.7 &   0.00 &   0.00 &  0.00 & 2.129 & 5.432 & 7.790 & 9.562\\
\hline

\end{tabular}
\end{table}
\newpage 

\newpage 

\begin{table}
\caption{$pp$ and $nn$ total cross sections in asymmetric nuclear matter
calculated with the DBHF model and according to Eq.(1).
Kinematics and definition of variables are given in the text. 
} 
\begin{center}
\begin{tabular}{|c||c|c|c|c|}
$\alpha$ & $q_0(MeV/c)$  & $k_F(fm^{-1})$ & $\sigma_{pp}(mb)$ & $\sigma _{nn}(mb)$ \\  
 \cline{1-5}
 0.2  &  100    & 0.2 & 154.6  & 160.0  \\ 
      &         & 0.4 & 131.2  & 135.9  \\ 
      &         & 0.5 & 110.5  & 104.9  \\ 
      &         & 0.6 & 78.42  & 47.76  \\ 
      &         & 0.7 & 32.71  & 0.0    \\ 
      &  200    & 0.2 & 34.38  & 35.63\\
      &         & 0.4 & 29.17  & 30.75\\
      &         & 0.6 & 24.44  & 25.62\\
      &         & 0.8 & 20.07  & 20.23\\
      &         & 1.0 & 15.76  & 14.28\\
      &         & 1.1 & 13.28  & 10.55\\
      &         & 1.2 & 10.55  & 6.567\\
      &         & 1.3 & 7.362  & 2.528\\
      &         & 1.4 & 4.405  & 0.0  \\
      &         & 1.5 & 1.660  & 0.0  \\
      &  300    & 0.2 & 21.41  & 22.04\\
      &         & 0.4 & 19.26  & 20.06\\
      &         & 0.6 & 17.56  & 18.21\\
      &         & 0.8 & 16.18  & 16.48\\
      &         & 1.0 & 14.97  & 14.80\\
      &         & 1.1 & 14.32  & 13.85\\
      &         & 1.2 & 13.63  & 12.86\\
      &         & 1.3 & 12.89  & 11.86\\
      &         & 1.4 & 12.24  & 10.81\\
      &         & 1.5 & 11.70  & 9.893\\
      &         & 1.6 & 11.22  & 8.877\\ 
      &         & 1.7 & 10.76  & 7.731\\
      &  400    & 0.2 & 19.37  & 19.60\\
      &         & 0.4 & 18.72  & 18.94\\
      &         & 0.6 & 18.29  & 18.37\\
      &         & 0.8 & 17.99  & 17.86\\
      &         & 1.0 & 17.71  & 17.34\\
      &         & 1.1 & 17.57  & 17.06\\
      &         & 1.2 & 17.48  & 16.81\\
      &         & 1.3 & 17.43  & 16.60 \\                 
      &         & 1.4 & 17.52  & 16.45\\
      &         & 1.5 & 17.75  & 16.44\\
      &         & 1.6 & 18.10  & 16.44\\
      &         & 1.7 & 18.46  & 16.36\\
 \cline{1-5}
 0.4  &  100    & 0.2 & 153.8  & 162.9  \\ 
      &         & 0.4 & 131.3  & 138.6  \\ 
      &         & 0.5 & 111.6  & 100.0  \\ 
      &         & 0.6 & 90.30  & 29.40  \\ 
      &         & 0.7 & 56.24  & 0.0    \\ 
      &         & 0.8 & 17.96  & 0.0    \\ 
      &  200    & 0.2 & 34.23  & 36.32\\
      &         & 0.4 & 28.94  & 31.65\\
      &         & 0.6 & 24.23  & 26.33\\
      &         & 0.8 & 20.10  & 20.36\\
      &         & 1.0 & 16.44  & 13.30\\
      &         & 1.1 & 14.39  & 9.038\\
      &         & 1.2 & 12.12  & 4.445\\
      &         & 1.3 & 9.586  & 0.0  \\
      &         & 1.4 & 7.127  & 0.0  \\
      &         & 1.5 & 4.657  & 0.0  \\
      &         & 1.6 & 2.404  & 0.0  \\
      &         & 1.7 & 0.467  & 0.0  \\
      &  300    & 0.2 & 21.32  & 22.39\\
      &         & 0.4 & 19.12  & 20.53\\
      &         & 0.6 & 17.41  & 18.61\\
      &         & 0.8 & 16.10  & 16.69\\
      &         & 1.0 & 15.09  & 14.73\\
      &         & 1.1 & 14.56  & 13.66\\
      &         & 1.2 & 14.00  & 12.52\\
      &         & 1.3 & 13.43  & 11.36\\
      &         & 1.4 & 12.97  & 10.09\\
      &         & 1.5 & 12.59  & 8.901\\
      &         & 1.6 & 12.38  & 7.679\\ 
      &         & 1.7 & 12.31  & 6.277\\
      &  400    & 0.2 & 19.33  & 19.73\\
      &         & 0.4 & 18.69  & 19.09\\
      &         & 0.6 & 18.30  & 18.46\\
      &         & 0.8 & 18.10  & 17.84\\
      &         & 1.0 & 17.94  & 17.19\\
      &         & 1.1 & 17.87  & 16.85\\
      &         & 1.2 & 17.85  & 16.52\\
      &         & 1.3 & 17.91  & 16.23\\
      &         & 1.4 & 18.12  & 15.94\\
      &         & 1.5 & 18.48  & 15.78\\
      &         & 1.6 & 19.02  & 15.66\\
      &         & 1.7 & 19.68  & 15.45\\
 \cline{1-5}
 0.6  &  100    & 0.2 & 149.5  & 165.8  \\ 
      &         & 0.4 & 126.6  & 141.1  \\ 
      &         & 0.5 & 112.7  & 92.79  \\ 
      &         & 0.6 & 95.88  & 10.11  \\ 
      &         & 0.7 & 75.14  & 0.0    \\ 
      &         & 0.8 & 49.49  & 0.0    \\ 
      &  200    & 0.2 & 33.21  & 37.04\\
      &         & 0.4 & 27.77  & 32.59\\
      &         & 0.6 & 23.45  & 27.08\\
      &         & 0.8 & 20.01  & 20.50\\
      &         & 1.0 & 17.14  & 12.35\\
      &         & 1.1 & 15.53  & 7.168\\
      &         & 1.2 & 13.70  & 2.156\\
      &         & 1.3 & 11.71  & 0.0  \\
      &         & 1.4 & 9.672  & 0.0  \\
      &         & 1.5 & 7.618  & 0.0  \\
      &         & 1.6 & 5.727  & 0.0  \\
      &         & 1.7 & 4.124  & 0.0  \\
      &  300    & 0.2 & 20.82  & 22.76\\
      &         & 0.4 & 18.57  & 21.03\\
      &         & 0.6 & 17.04  & 19.05\\
      &         & 0.8 & 16.00  & 16.94\\
      &         & 1.0 & 15.24  & 14.72\\
      &         & 1.1 & 14.84  & 13.46\\
      &         & 1.2 & 14.42  & 12.20\\
      &         & 1.3 & 14.02  & 10.82\\
      &         & 1.4 & 13.70  & 9.389\\
      &         & 1.5 & 13.51  & 8.046\\
      &         & 1.6 & 13.54  & 6.488\\ 
      &         & 1.7 & 13.77  & 4.664\\
      &  400    & 0.2 & 19.15  & 19.87\\
      &         & 0.4 & 18.57  & 19.27\\
      &         & 0.6 & 18.33  & 18.58\\
      &         & 0.8 & 18.26  & 17.86\\
      &         & 1.0 & 18.21  & 17.08\\
      &         & 1.1 & 18.21  & 16.65\\
      &         & 1.2 & 18.27  & 16.26\\
      &         & 1.3 & 18.43  & 15.85\\
      &         & 1.4 & 18.75  & 15.48\\
      &         & 1.5 & 19.26  & 15.22\\
      &         & 1.6 & 20.00  & 14.91\\
      &         & 1.7 & 20.91  & 14.50\\
\end{tabular}
\end{center}
\end{table}


\begin{references}
\bibitem{SK05} F. Sammarruca and P. Krastev, nucl-th/0506081.
\bibitem{SM} F. Sammarruca, X. Meng, and E.J. Stephenson, Phys. Rev. C {\bf 62},          
014614 (2000). 
\bibitem{hh1} T. Alm, G. Ropke, and M. Schmidt, Phys. Rev. C {\bf 50}, 31 (1994). 
\bibitem{hh2} A. Bohnet, N. Ohtsuka, J. Aichelin, R. Linden, and A. Faessler, 
Nucl. Phys. {\bf A494}, 349 (1989). 
\bibitem{SBK} F. Sammarruca, W. Barredo, and P. Krastev, Phys. Rev. C {\bf 71},
064306 (2005). 
\bibitem{MSU} B.-A. Li, P. Danielewicz, and W. Lynch, nucl-th/0503038. 
\bibitem{Li05} B.-A. Li and Lie-Wen Chen, nucl-th/0508024. 
\bibitem{QLi04} Q. Li, Z. Li, and E. Zhao, Phys. Rev. C {\bf 69}, 017601 (2004).         

\end{references}
\end{document}